\documentstyle[preprint,aps]{revtex}
\begin{document}
\draft
\title{Quantum-Phase Transitions of Interacting Bosons\\
	and the Supersolid Phase}
\author{Anne van Otterlo$^1$, Karl-Heinz Wagenblast$^2$, Reinhard Baltin$^2$,\\
	C. Bruder$^2$, Rosario Fazio$^2$, and Gerd Sch\"{o}n$^2$}
\address{$^1$Theoretische Physik, ETH-H\"{o}nggerberg,
	CH-8093 Z\"{u}rich, Switzerland}
\address{$^2$Institut f\"ur Theoretische Festk\"orperphysik,
	Universit\"at Karlsruhe, D-76128 Karlsruhe, FRG}
\maketitle

\begin{abstract}
We investigate the properties of strongly interacting bosons in
two dimensions at zero temperature using mean-field theory,
a variational Ansatz for the ground state wave function, and
Monte Carlo methods. With on-site and short-range interactions
a rich phase diagram is obtained. Apart from the homogeneous
superfluid and Mott-insulating phases, inhomogeneous
charge-density wave phases appear, that are stabilized by the
finite-range interaction. Furthermore, our analysis demonstrates the
existence of a supersolid phase, in which both long-range order
(related to the charge-density wave) and off-diagonal long-range
order coexist. We also obtain the critical exponents for the
various phase transitions.
\end{abstract}


\section{Introduction}
In recent years strongly interacting bosons attracted a lot of
attention. Experiments on thin granular films show a zero temperature
superconductor to insulator (SI) phase transition as a function of the film
thickness~\cite{gold} or magnetic field~\cite{hpsi}, and experiments on
Josephson junction arrays (JJA) exhibit the same SI transition as a function
of the ratio between Josephson coupling and charging energy~\cite{jja}. These
findings are interpreted in terms of a bosonic model~\cite{efe,doni,fwgf} in
which the competition between the hopping of Cooper pairs and the Coulomb
interaction between them is responsible for the SI transition.
In relation to the effect of disorder on the SI transition a Bose-glass phase
is discussed in Refs.~\cite{fwgf,f-db,zimrev}.

An interesting extension of the minimal model with on-site interaction only is
to account for finite-range interactions, which are present in Josephson
junction arrays and $^4$He-films on substrates in two dimensions, and bulk
$^4$He in three dimensions. Artificially fabricated JJA's are ideal model
systems to study
the superconductor to Mott-insulator transition as a function of the coupling
constants or magnetic field~\cite{jja}. The finite range of the interaction in
these systems leads to commensurability and frustration effects.
Furthermore, coexistence phases may appear that are called supersolids. In
these phases a charge-density wave (CDW) that is stabilized by the interaction
may coexist with superfluidity. In other words the system has diagonal
long-range order (LRO) and off-diagonal LRO at the same time.

The supersolid phase was studied in the early seventies after Andreev and
Lifshitz~\cite{andlif} suggested that vacancies in a quantum crystal such as
$^4$He might Bose-Einstein condense at low temperatures without destroying the
crystal structure, thereby establishing a superfluid solid~\cite{l-ss}, or
supersolid. Normally bosons at zero temperature are either superfluid (with
off-diagonal LRO) or solid (with diagonal LRO). However, for a finite
range of the interactions between the bosons a coexistence phase was
predicted within mean-field approximations~\cite{mt-ss,lf-ss,bfs,rs-ss}.
Experiments have been performed on $^{4}$He, but no positive
identification of this coexistence phase has yet been made. There are,
however, hints towards such a phase \cite{m-ss,lg-ss}.

Recently several other kinds of coexistence phases were studied.
The possibility of a spontaneous vortex anti-vortex lattice in
superfluid films was explored in Ref.~\cite{vavl} and a coexistence
phase of superfluidity and hexatic orientational order was proposed
in Ref.~\cite{hex}. Orientational order in incompressible quantum Hall fluids
is discussed in Ref.~\cite{leon}. Collinear supersolids were studied in
Refs.~\cite{gergy,rs-mc}. Finally, we mention the relation between 2D bosons
and 3D flux-lines in type II superconductors (high $T_{c}$ materials)
in a magnetic field~\cite{nelson,fl-d,feigel}. Also in these
systems different kinds of LRO may coexist and the equivalent of
the supersolid is discussed in Refs.~\cite{blatt,frey}.
Related is the question whether or not vortices may form a disentangled
liquid, which would imply a normal ground state for bosons with
long-range Coulomb interaction~\cite{feigel}.

In this paper both the SI transition and the supersolid phase will be
studied in some detail. In the next section several related models for
interacting bosons will be discussed, of which the Bose-Hubbard model
is the most general. In the limit of infinite on-site interaction the
Bose-Hubbard model reduces to the spin $\frac{1}{2}$ XXZ model.
For a large number of bosons per lattice site the Quantum-Phase model
is applicable. The phase diagram of the latter two models are first
determined within mean-field theory in Section~\ref{sec-mf}.
A variational Ansatz for the ground state
wave function that treats all three models on equal footing~\cite{baltin}
is presented in Section~\ref{sec-vm}. In Section \ref{sec-nm} more
accurate results for the phase diagram will be obtained by means of
Quantum Monte Carlo. Part of the data has already been
published as Ref.~\cite{ow-ss}. Finally the paper closes with a
discussion of the various results in Section~\ref{sec-d}.

\section{Three model Hamiltonians}
\label{sec-tmh}
A convenient starting point for the description of interacting lattice
bosons is the Bose-Hubbard Hamiltonian
\begin{equation}
	H_{BH}=\frac{1}{2}\sum_{ij}n_{i}U_{ij}n_{j} -\mu\sum_{i}n_{i}
	-\frac{t}{2}\sum_{\langle ij\rangle}(b^{\dagger}_{i}b_{j}+
b^{\dagger}_{j}b_{i})\;,
\label{bh}
\end{equation}
where $b^{\dagger}, b$ are the creation and annihilation operators
for bosons that satisfy the commutation relation
$[b_i,b_j^{\dagger}]=\delta_{ij}$ and $n_i=b^{\dagger}_ib_i$ is the number
operator. The first term describes
the density-density interaction between the bosons. We will take it
to be short range, $U_{0}$ for on-site, $U_{1}$ for nearest neighbors,
and $U_{2}$ for second-nearest neighbors. The second term describes
the coupling of the density to a chemical potential, and the third
describes the hopping from site to site with hopping integral $t$. The
first two terms will cause the bosons to be localized if the interaction
dominates over the hopping. In the opposite case where the hopping
dominates the bosons will form a superfluid.

In the limit of large on-site interaction $U_{0}$, only the states with
0 and 1 boson per lattice site survive, and by identifying $b^{\dagger}
= S^{+} = S^{x}+iS^{y}$ and $n = S^{z}+\frac{1}{2}$ the mapping to the
spin $\frac{1}{2}$ XXZ Hamiltonian is obtained. It reads
\begin{equation}
	H_{XXZ}=\frac{1}{2}\sum_{i,j}S^{z}_{i}U_{ij}S^{z}_{j}
	-h\sum_{i}S^{z}_{i}
	-t\sum_{\langle ij\rangle}[S^{x}_{i}S^{x}_{j}+S^{y}_{i}S^{y}_{j}]\;,
\label{xxz}
\end{equation}
where $h=\mu-\frac{1}{2}\sum_i U_{0i}$.
This Hamiltonian has been
investigated extensively on the mean-field level for nearest-neighbor
as well as next nearest-neighbor interactions.

In the limit of a large number of bosons per lattice site one may
parameterize $b^{\dagger}=\sqrt{\rho}e^{i\varphi}$ and take $\rho$
equal to the boson density. Thus, only the phase remains as a dynamic
variable and the Quantum-Phase Hamiltonian reads
\begin{equation}
	H_{QP}=\frac{1}{2}\sum_{i,j}n_{i}U_{ij}n_{j} -\mu\sum_{i}n_{i}
	-J\sum_{\langle ij\rangle}\cos(\varphi_{i}-\varphi_{j})\;,
\label{qp}
\end{equation}
where $J=\rho t$. The phase and number are canonical conjugate variables
and satisfy $[\varphi_i,n_j]=i\delta_{ij}$.
This is exactly the Hamiltonian for Josephson junction arrays with
an applied gate voltage in order to tune the chemical potential
\cite{bfkos}. It is convenient to introduce a parameter $n_0=
\mu/\sum_{i}U_{ij}$, which allows us to rewrite the Coulomb interaction
and chemical potential term in Eqs.~(\ref{bh}) and (\ref{qp}) as
\begin{equation}
	H_{\mbox{\footnotesize int}}=
	\frac{1}{2}\sum_{i,j}(n_{i}-n_{0})U_{ij}(n_{j}-n_{0}) \;.
\label{int}
\end{equation}
We expect that the properties of the Quantum-Phase model will be periodic in
the variable $n_{0}$ with period 1.

\section{Mean-field}
\label{sec-mf}
A first insight into the properties of these models may be gained
by considering their mean-field solutions. The mean-field approximation
for the spin model in Eq.~(\ref{xxz}) is straightforward and was carried
out by various authors \cite{mt-ss,lf-ss,bfs}. The question of a
supersolid phase in this model was first addressed in Ref.~\cite{lf-ss}
and investigated further in Ref.~\cite{bfs}.
We review the spin mean-field theory in Subsect.~\ref{subsec-smf}.
A mean-field approximation for the Quantum-Phase model which is able
to identify both diagonal LRO and off-diagonal LRO appeared in
Ref.~\cite{rs-ss}. It is reviewed in Subsect.~\ref{subsec-qpm}.
Mean-field approaches to the Bose-Hubbard Model have been discussed
in Refs.~\cite{fwgf,skpr}, but to our present knowledge there exists no
mean-field approximation for the Bose-Hubbard model which allows for the
identification of a supersolid phase.

In Sect.~\ref{sec-vm} we present a variational Ansatz for the many-body wave
function that allows the three models to be treated on an equal footing. This
is important, as it allows for a comparison between the models on the same
level of approximation. Since we use a factorizable Ansatz for the
wave function, it also has mean-field character.

The validity of the mean-field approximation in two dimensions is often
limited, since the lower critical dimension for models with continuous
symmetries is two, which is the dimension we are particularly interested
in. This is a serious problem at finite temperature where the models do not
even exhibit superfluid long-range order as formulated in the theorem of
Mermin and Wagner, and Hohenberg~\cite{rmwh}.
Useful results may, however, be gained from mean-field approximations at zero
temperature where quantum dynamics formally adds an extra dimension, the
imaginary time axis. The systems then do have long-range order and mean-field
results may be qualitatively correct. This will be shown in the present
article by scaling arguments and confirmed by the Monte Carlo simulations. In
the remainder of this section we present the different mean-field type
approaches.

\subsection{Spin Mean-Field}
\label{subsec-smf}
The spin mean-field theory can be formulated by linearizing the spin-spin
interaction. This yields the standard self-consistency equations for the
magnetization
\begin{equation}
	\langle{\bf S}_{i}\rangle=-\frac{1}{2}
	\frac{{\bf H}_{i}}{\mid{\bf H}_{i}\mid}
	\tanh\left(\frac{\beta}{2}\mid{\bf H}_{i}\mid\right)\;,
\end{equation}
where the subscript $i$ refers to the sublattice and $\beta$ is the inverse
temperature (we use units in which $k_{B}=\hbar=c=1$). In order to identify
spatially varying solutions one needs to introduce several sublattices,
at least three for interactions including nearest and next-nearest
neighbors. The symmetry in the $xy$-plane reduces the number of
independent variables and most of the phase boundaries can be
determined analytically. If we allow for three different sublattices
($A$, $B$ and $C$), the effective field is given by
\begin{equation}
	{\bf H}_{A}=(-4t\langle S^{x}_{B}\rangle,
	-4t\langle S^{y}_{B}\rangle,2U_{1}\langle S^{z}_{B}\rangle
	+2U_{2}\langle S^{z}_{C}\rangle-h) \;,
\end{equation}
and corresponding equations for ${\bf H}_{B}$ and ${\bf H}_{C}$.

The connection to hard-core bosons is made by remembering that $S^z$
corresponds to the particle number, and $S^{x/y}$ to the boson creation
operators. A staggered magnetization in $z$-direction implies a solid (
charge-density wave) ordering of the bosons.
Finite magnetization in $xy$-direction
translates into off-diagonal LRO for the boson model. The coexistence phase of
both types of LRO is the supersolid phase for the bosons. The result is shown
in Fig.~\ref{spinfig} for $U_{2}=0.1\;U_{1}$. Two different types of solid
ordering occur for different chemical potential, i.e., $(\pi,\pi)$ ordering
around half-filling, and additional $(\pi,0)$ and
$(0,\pi)$ ordering around quarter filling.

\subsection{Quantum-Phase Model}
\label{subsec-qpm}
Now we turn to the mean-field treatment of soft-core bosons following
Refs.~\cite{rs-ss,khw-d}. After decoupling the hopping and the interactions
between the different sites in Eq.~(\ref{qp}), one arrives at a local problem
\begin{equation}
	H_{mf}^i\;|\psi_i\rangle = E_i\;|\psi_i\rangle \; ,
\end{equation}
with the $2\pi$-periodic solutions for the wave function in the phase
representation
\begin{equation}
  	\psi_i(\varphi_i) = \exp(i\eta_i x_i) \; f_i(x_i) \; ,
\label{eq:sol}
\end{equation}
where $\varphi_i=2\,x_i$, $\eta_i= 2\,n_{0} \;-\; \frac{2}{U_0}
\sum_{j \neq i} U_{ij}(\langle n_{j}\rangle-n_{0})$.
After these substitutions the following Mathieu equation for $f$ is obtained
\begin{equation}
	\left\{ \frac{\partial^2}{\partial x_i^2} \;
	+\;e_i\;+\;2\,r_i\cos(2\,x_i) \right\} f_i(x_i) \;=\;0
\end{equation}
\begin{equation}
	\mbox{with}\ \ \  r_i\;=\;\frac{4J}{U_0}
	\sum_{\beta}\langle\cos(\varphi_{i+\beta})\rangle \; .
\end{equation}
The sum runs over nearest neighbors $\beta$ of $i$. The parameter
$e_i$ is proportional to the energy and should be minimized. The ground
state is defined via the self-consistency equation
\begin{eqnarray}\nonumber
	\langle n_i\rangle\ \ \ \ &=&\ \ \ \langle\psi_i^0\,|\;n_i\;
|\psi_i^0\rangle\\
	\langle\cos(\varphi_i)\rangle&=&\langle\psi_i^0\,|\,
\cos(\varphi_i)\,|\psi_i^0\rangle \;.
\label{sc}
\end{eqnarray}
In addition the free energy has to be minimized in order to have an
unambiguous solution of Eq.~(\ref{sc}). The numerical solution of this set of
equations is straightforward. In order to allow for spatially varying solutions
we again define sublattices. Our numerical solutions confirm the results of
Ref.~\cite{rs-ss}, however, some details differ.

For on-site interaction we obtain Mott insulating lobes centered around
integer values $N$ of $n_0$, see Fig.~\ref{mottfig}. In these incompressible
lobes the density is pinned to $N$. At half integer values of $n_0$ the
critical value of the hopping vanishes, due to the degeneracy in energy of
states with $N$ and $N+1$ particles per site.

Including nearest-neighbor interaction the calculation has to be performed on
two sublattices. We obtain two different kinds of insulating lobes, one with
integer filling and homogeneous density, another with half-integer filling and
a checkerboard charge-density wave, or solid order, centered around half
integer values of $n_0$. In addition we find a region of supersolid phase
around the half lobe where the checkerboard charge-density wave and
superfluidity coexist, see Fig.~\ref{cbssfig}. Again, the superfluid phase
occupies the parts on the right of the phase diagram and also between the
insulating lobes down to zero hopping, in slight contrast to the result of
Ref.~\cite{rs-ss}.

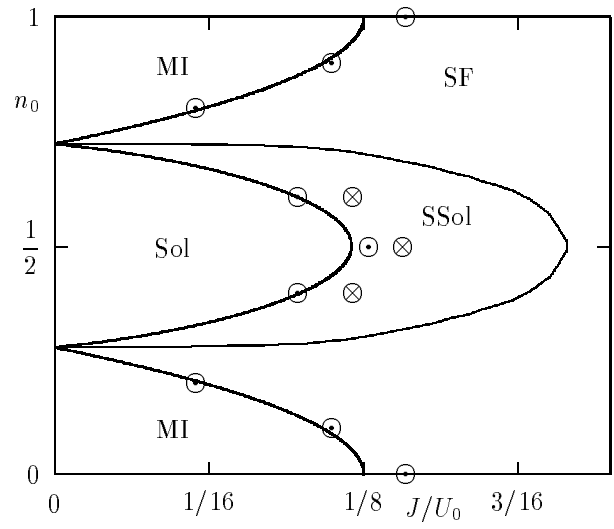
\begin{figure}
\caption{: Supersolid region ``SSol'' at $n_{0}$= 0.5 as a function of
	$U_1/U_0$ in the mean-field approximation of
	Section~\protect\ref{subsec-qpm}. Inset: Occupation-number probability
	$|c_{2}|^{2}$ at $n_{0}$= 0.5 for the two sublattices $A$ and
	$B$ at the particular value of $U_1/U_0=0.2$.}
\label{numbfig}\end{figure}

The fraction of supersolid phase in the phase diagram depends on the ratio
$U_{1}/U_{0}$, see Fig.~\ref{numbfig}. In the hard-core limit $U_0 \rightarrow
\infty$ the supersolid phase vanishes. An expansion of the ground state wave
function $|\psi_i^0\rangle=\sum_{n}c^{i}_{n}|n\rangle$ into particle number
eigenstates $|n\rangle$ yields more information about the nature of the
supersolid phase. The inset of Fig.~\ref{numbfig} shows $|c_{2}|^{2}$ at
$n_{0}$=0.5 as a function of $J/U_{0}$. The figure indicates that double
occupancy is important for the supersolid phase. Thus, we conclude that the
supersolid phase is favored by fluctuations in the particle number.

The inclusion of further interactions leads to more structure in the phase
diagram. For next-nearest neighbor interaction we introduce three
sublattices. As a result, The phase diagram shows a lobe with quarter filling
and a phase where this quarter structure and superconductivity coexist,
see Fig.~\ref{nnnfig}.

\subsection{Hard-core vs. Soft-core}
\label{subsec-hcsc}
The previous subsections showed two extreme cases of the Bose-Hubbard model,
the hard-core limit and the limit of large particle numbers. Although the
phase diagrams look similar, these two limits behave qualitatively different
as far as the supersolid phase is concerned.

In the hard-core limit the existence of the supersolid is related to a finite
next-nearest neighbor interaction. The supersolid does not exist in the model
including nearest-neighbor interaction only~\cite{bfs}. Furthermore, there is
no supersolid phase at exactly half filling. This leads to the following
interpretation of the supersolid phase. At densities corresponding to half
filling the particles form an incompressible solid. Away from half filling
superfluidity is enabled by defects in the solid structure that Bose-Einstein
condense. This interpretation does, however, not explain why next-nearest
neighbor interaction is necessary for the formation of the supersolid.

In the limit of many particles per site the Quantum-Phase model applies. In
this case the supersolid phase already exists for nearest-neighbor interaction
only and at half filling. This is related to excitations which are forbidden
in the hard-core limit. We observe that a large nearest-neighbor interaction
$U_{1}$ or small on-site interaction $U_{0}$ favors the supersolid, in the
hard-core limit the supersolid is suppressed. In the soft-core limit the
supersolid phase is also present at half-filling at the tip of the
checkerboard lobe. Thus, it seems that the system itself generates the defects
(particle-hole excitations) that Bose condense, thereby turning the solid into
the supersolid.

\section{Variational Method}
\label{sec-vm}
Here we present a method which allows us to treat the three models within one
scheme~\cite{baltin}. The idea is to guess a variational wave function for the
problem \cite{gutz}. We use a variational wave function that is inspired by
earlier successful Gutzwiller-like treatments of the spin model~\cite{huse}
and the Bose-Hubbard model~\cite{roko,kraut}. The occupation number
representation is well suited for all three models. Explicitly we resolve the
identity as
\begin{equation}
	1=\sum_{\{n_i\}}|\{n_i\}\rangle\langle\{n_i\}|
\end{equation}
and express the trace as
\begin{equation}
	\mbox{Tr ...}=\sum_{\{n_i\}}\langle\{n_i\}| ... |\{n_i\}\rangle,
\end{equation}
where the $n_i$ are in $[0,1]$ for the spin $\frac{1}{2}$ XXZ model,
in $[0..\infty]$ for the Bose-Hubbard (BH) model, and in
$[-\infty..\infty]$ for the
Quantum-Phase (QP) model. Besides the difference in the allowed values of
$n_i$, the matrix elements differ for the three models, e.g.,
\begin{eqnarray}\nonumber
	\mbox{BH model:}\;\;\;\;\;\; & & \langle\{n_i'\}|b_k|\{n_i\}\rangle
        	= \langle\{n_{i\neq k}'\}|\{n_{i\neq k}\rangle
                       \langle n_k'|n_k-1\rangle \sqrt{n_k} \\
	\mbox{QP model:}\;\;\;\;\;\; & &
		\langle\{n_i'\}|\exp{i\varphi_k}|\{n_i\}\rangle
        	= \langle\{n_{i\neq k}'\}|\{n_{i\neq k}\rangle
                       \langle n_k'|n_k-1\rangle
\end{eqnarray}
As our variational wave function we take the product of single-site wave
functions. This amounts to neglecting certain correlations and is therefore
similar to a mean-field approximation. We use the following Ansatz
$$
	|\psi\rangle=\prod_{i=1}^{N}\frac{1}{\sqrt{Z_{i}}}
	\sum_{\{n_{i}\}}\exp\left\{-k_{i}
	(n_{i}-m_{i})^{2}/2\right\}|n_{i}\rangle
$$
\begin{equation}
	\mbox{where }\ \ \ \ Z_{i}=
	\sum_{\{n_{i}\}}\exp\left\{-k_{i}(n_{i}-m_{i})^{2}\right\} \;,
\label{ansatz}
\end{equation}
and $k_{i}$ and $m_{i}$ are real variational parameters. The single-site wave
function is a superposition of boson number eigenstates weighted with a
Gaussian. The average is controlled by $m_i$, the width by $k_i$, and spatial
variations (different sublattices $i$) are allowed.
By minimizing the energy expectation value $E=\langle\psi|H|\psi\rangle$ with
respect to $k_{i}$ and $m_{i}$ we obtain the ground state wave function within
our approximation. In the following, expectation values are understood
to be calculated with the ground state wave function. As the discrete
sums can in general not be performed analytically, we treat the problem
numerically. For on-site and nearest-neighbor interactions two sublattices
arranged in a checkerboard configuration
are introduced. We define the superconducting order parameter as
$\langle\exp(i\varphi)\rangle$ for the QP model, $\langle b\rangle$ for the BH
model, and $\langle S^{-} \rangle$ for the XXZ model. The structure factor
\begin{eqnarray}\label{eqstr}
	S(\vec{q}) & = & \bigg(\frac{1}{N}\bigg)^{2}\sum_{i,j}
	\exp[i\vec{q}\cdot(\vec{r}_{i}-\vec{r}_{j})]\langle n_{i}n_{j}\rangle
\end{eqnarray}
yields information about the solid order. A non-vanishing structure factor at
finite $\vec{q}$ signals diagonal LRO. By decomposing the lattice into two
sublattices we gain information about $S(\pi,\pi)$. A finite $S(\pi,\pi)$
corresponds to a checkerboard arrangement of the particles. For next-nearest
neighbor interactions we introduce four sublattices which yields additional
information about $S(\pi,0)$ and $S(0,\pi)$.

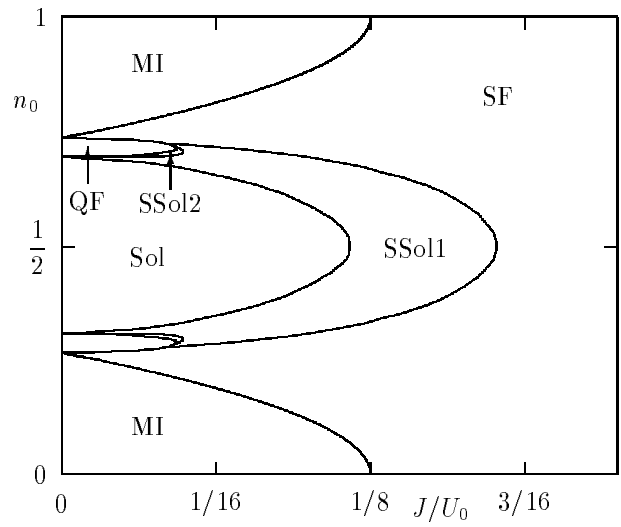
\begin{figure}\caption{: Phase diagram for the Quantum-Phase model as
	determined from the variational Ansatz with nearest-neighbor
	interactions, $U_1/U_0=1/5$.}
\label{qpvarfig}\end{figure}

Figure~\ref{qpvarfig} shows the phase diagram for the QP model with
on-site and nearest-neighbor interaction obtained with our
variational Ansatz. Both the superfluid-insulator and the crystalline order
transition can be identified and agree with previous mean-field calculations
\cite{rs-ss,ow-ss}. The phase boundaries are periodic in $n_{0}$. We
find that superfluidity sets in simultaneously on both sublattices $A$
and $B$. Within the supersolid phase the value of the superconducting
order parameter on $A$ differs from that on $B$, but both are nonzero.
Beyond the crystalline order phase boundary the order parameter
does not exhibit any difference on the sublattices.

For the XXZ model the variational method confirms that there is no
supersolid for on-site and nearest-neighbor interaction only.
In the presence of next-nearest neighbor interaction we find
supersolid phases in perfect agreement with \cite{bfs} and Fig.~\ref{spinfig}.

The phase diagram for the BH Hamiltonian is shown in Fig.~\ref{bhvarfig}.
The size of the lobes decreases with increasing $n_0$. At point $\alpha$
the supersolid vanishes. This might be due to the lower bound for the
occupation numbers $n\geq 0$. For small $n_0$ charge fluctuations are
suppressed. Hence, we conclude that charge fluctuations are necessary for the
supersolid. For large $n_{0}$ the phase boundaries of the QP model approach
those of the BH model, if the hopping is rescaled according to
Section~\ref{sec-tmh}, i.e., if $t$ and $J/n_{0}$ are identified.

\section{Quantum Monte Carlo}
\label{sec-nm}
Several methods are available for determining the phase diagram for
the three model Hamiltonians by numerical means. Exact diagonalization
is possible for the XXZ model and small systems, i.e., less than about 26
lattice sites. Furthermore it is possible to perform Monte Carlo
simulations for all the three models on much larger systems.
The Bose-Hubbard model was studied in one dimension in
Refs.~\cite{bsz,nfsb}, and one data point for the phase boundary was
obtained in two dimensions in Ref.~\cite{kt-mc}.
The Bose-Hubbard model was more recently studied in the context of
a supersolid phase in Ref.~\cite{gergy}.
The two-dimensional Quantum-Phase model was studied in Ref.~\cite{swgy}
in relation to the universal conductivity at the SI transition and in
Refs.~\cite{rs-mc,ow-ss} in relation to supersolid phases.

In this section we present more results on Monte Carlo simulations
of the two-dimensional QP model in addition to those in Ref.~\cite{ow-ss}.
We discuss the on-site problem and compare to the analytic results of
$t/U$ expansions \cite{fm-bh,fm-ex}. The inclusion of nearest-neighbor
interactions allows us to study the supersolid phase.

\subsection{Duality Transformation}
\label{subsec-dt}
In order to study the Quantum-Phase model by means of Monte Carlo
we map the 2D quantum model onto a (2+1)D classical model of discrete
divergence-free currents. The essential feature of this mapping were
presented in
Ref.~\cite{fl-d} and the derivation makes use of duality
transformations that were developed in Ref.~\cite{jkkn}. We start
from the basic expression for the partition function $Z$ for the
Quantum-Phase model of Eq.~(\ref{qp})
\begin{equation}
	Z=\mbox{Tr}\;\exp(-\beta H_{QP})\;,
\label{az}
\end{equation}
where $\beta$ is the inverse temperature. We go over to a Euclidean
path-integral formulation by introducing time-slices, i.e., dividing
$\beta$ in $N_{\tau}$ intervals of size $\epsilon$, such that
$N_{\tau}\epsilon=\beta$. Inserting complete sets of states at each
time slice we arrive at
$$
	Z=\sum_{\{n_{i,\tau}=0,\pm1,\pm2,\cdots\}}
	\int{\cal D}\varphi_{i,\tau}\exp{\Big\{
	-\frac{\epsilon}{2}}\sum_{ij,\tau}
	n_{i,\tau}U_{ij}n_{j,\tau}+ \epsilon\mu\sum_{i,\tau}n_{i,\tau}
$$
\begin{equation}
	+i\sum_{i,\tau}n_{i,\tau}\dot{\varphi}_{i,\tau}+
	\epsilon J\sum_{\langle ij\rangle,\tau}
	\cos(\varphi_{i,\tau}-\varphi_{j,\tau}-A_{ij}){\Big\}}\;.
\label{laz1}
\end{equation}
We included a coupling to a vector potential $A_{ij}=(2\pi/\Phi_{0})
\int^{j}_{i}{\bf A}\cdot d{\bf l}$ for later convenience.
In order to proceed, we make one approximation, the Villain
approximation~\cite{vill} for the cosine term. This amounts to
expanding the cosine around all its minima
\begin{equation}
	\exp\{\epsilon J\cos(\varphi_{i\tau}-\varphi_{j\tau}-A_{ij})\}
	\approx\!\!\!\sum_{\{m_{ij,\tau}\}}
	\exp\{-\frac{\epsilon J F(\epsilon J)} {2}
	(\varphi_{i\tau}-\varphi_{j\tau}-2\pi m_{ij,\tau}-A_{ij} )^{2}\}\;,
\label{eq:av}
\end{equation}
where $m$ is a directed discrete field that lives on the bonds between
lattice sites. The function $F$ is determined by demanding that the
first two Fourier coefficients of the expressions on the left and
right hand side of Eq. (\ref{eq:av}) are equal~\cite{vill} and is given by
\begin{equation}
	F(x)=\frac{1}{2x\log\{I_{0}(x)/I_{1}(x)\}}\;.
\label{vill}
\end{equation}
For large arguments $F$ approaches 1, as is clear from a direct
expansion of the cosine potential. After a subsequent Poisson
resummation, i.e., writing
$$
	\sum_{\{m_{ij,\tau}\}}f[m_{ij,\tau}]=
	\sum_{\{{\cal J}_{ij,\tau}\}}\int {\cal D}m f[m_{ij,\tau}]
	\exp(2\pi i\sum_{ij,\tau}m_{ij,\tau}{\cal J}_{ij,\tau})\;,
$$
an integration over the fields $m_{ij,\tau}$ and the phases
$\varphi_{i,\tau}$ yields a representation in terms of divergence-free
discrete current loops
$$
	Z=\left[\epsilon JF(\epsilon J)\right]^{-1}
	\sum_{\{{\cal J}^{\nu}_{i,\tau}\}}
	\delta(\nabla_{\nu}{\cal J}^{\nu})\exp\{-S[{\cal J}]\}\;,
$$
$$
	S[{\cal J}]=\frac{\epsilon}{2}\sum_{ij,\tau}
	{\cal J}^{\tau}_{i,\tau} U_{ij} {\cal J}^{\tau}_{j,\tau}
	-\epsilon\mu\sum_{i,\tau}{\cal J}^{\tau}_{i,\tau}
	+\frac{1}{2\epsilon J F(\epsilon J)}\sum_{i,\tau,a=x,y}
	\left({\cal J}^{a}_{i,\tau}\right)^{2}
	-i\sum_{i,\tau}\vec{A}_{i}\cdot\vec{{\cal J}}_{i,\tau}
$$
\begin{equation}
	=\sqrt{\frac{2}{K}}\left\{\sum_{ij\tau}({\cal J}^{\tau}_{i,\tau}-n_{0})
	\left(\delta_{ij}+\frac{U_{1}}{U_{0}}\delta_{\langle ij\rangle}\right)
	({\cal J}^{\tau}_{j,\tau}-n_{0})+\!\!\sum_{i\tau,a=x,y}\!\!
	\left({\cal J}^{a}_{i,\tau}\right)^{2}\right\}
	-i\sum_{i,\tau}\vec{A_i}\cdot\vec{{\cal J}}_{i,\tau}\;,
\label{al}
\end{equation}
where $\nu=x,y,\tau$. The effective coupling constant is given by
$\sqrt{2/K}$ with $K=8F(\epsilon J)J/U_{0}$, where $\epsilon$ is
defined implicitely by $\epsilon=1/\sqrt{U_0F(\epsilon J) J}$.
The time components ${\cal J}^{\tau}_{i,\tau}$ of the current correspond
to the boson numbers $n_{i,\tau}$
along their world lines. We keep track of the energy dependence of
the determinant that arises from integrating out the $m_{ij,\tau}$.
This is relevant for determining the energy density and specific heat
by means of Monte Carlo.
For zero magnetic field ($\vec{A}=0$) the action in the representation
of Eq.~(\ref{al}) is real, and therefore suitable for Monte Carlo
methods. Using the standard Metropolis algorithm we generate
configurations of currents in a system of size $L\times L\times
L_{\tau}$.
The condition that the currents ${\cal J}^{\nu}$ be divergence-free is taken
into account by making Monte Carlo steps that preserve the property
$\nabla_{\nu}{\cal J}^{\nu}=0$. Thus, we create or annihilate small current
loops at every lattice site in all three directions, as well as
periodic current loops that go through the whole system which is
taken to have periodic boundary conditions in all three directions.
The generation of configurations may be done in a canonical as well as
in a grand-canonical way. Here we work in the grand-canonical ensemble
in order to make contact to the mean-field phase diagrams.

In the last line of Eq.~(\ref{al}) we used
$\epsilon=1/\sqrt{U_0F(\epsilon J) J}$.
With this choice the couplings are isotropic for on-site interaction,
leading to an efficient numerical algorithm. The choice of the time lattice
spacing $\epsilon$ needs some justification. Introducing time slices
in the partition function Eq.~(\ref{laz1}) is exact in the limit
$\epsilon \rightarrow 0$. This would produce extremely anisotropic
couplings in the current-loop model and the numerical simulation
would become impossible. Thus, a more reasonable choice for the time
lattice spacing is necessary. The plasma frequency
$\omega_{p}=\sqrt{JU_0}$ is a natural frequency for spin waves.
A cutoff beyond this frequency by introducing a time spacing
$\epsilon \approx 1/\omega_{p}$ should not do any harm. We tested
the dependence on the choice of $\epsilon$ for the quantities of
interest and observed that they are not sensitive to variations of
$\epsilon$ by one order of magnitude, provided that the temperature
is kept constant, i.e., $\beta=1/T=const.=N_{\tau}\epsilon$.
Additional justification for the choice of a finite time spacing
$\epsilon$ in the present paper is given by the fact that we are
only interested in the critical regime, where a high-frequency cutoff
should be irrelevant.

\subsection{Finite-Size Scaling}
The mapping to the current-loop model Eq.~(\ref{al}) is well suited
for simulating the behavior of the Quantum-Phase model with short
range interactions. In practice we are able to study system sizes
up to $ 12 \times 12 \times 12$, or $10 \times 10 \times 25$,
respectively. The factor that limits the largest systems to have $L$= 12 is
the necessity to make the periodic current loops in the spatial
directions. Only they can change the superfluid stiffness, which is a
topological quantity that cannot be changed by making local current
loops. As phase transitions take place only in the thermodynamic
limit, we need an additional ingredient that relates the data obtained
on finite-size systems to infinite system size properties as the
critical coupling and exponents. This is provided by finite-size scaling.

Let us first consider the onset of superfluidity in the system.
Therefore we study the behavior of the superfluid stiffness $\rho_0$,
a quantity which measures the response to a twist in boundary
conditions for the phase. A finite value of $\rho_0$ in the thermodynamic limit
reflects long-range phase coherence, i.e., off-diagonal LRO or
superfluidity. A twist in the boundary conditions for the phases by an
angle $\Theta$ increases the free energy density by
\begin{equation}
	\Delta f=\frac{1}{2}\rho_0\left(\frac{\Theta}{L}\right)^2
\label{df}
\end{equation}
The twist may be realized by applying a magnetic field. In general
the frequency and wave-vector dependent stiffness $\rho(k,\omega_{\mu})$
can be defined as
\begin{equation}
	\rho(k,\omega_{\mu})=\left(\frac{1}{2e}\right)^{2}
	\sum_{i,\tau}\frac{-\delta^{2}\ln Z}
	{\delta A^{x}_{i,\tau}\delta A^{x}_{0}} \Bigg|_{\vec{A}=0}
  	e^{-i\omega_{\nu}\tau-ikr_{i}}\;.
\end{equation}
The zero-frequency and zero wave-vector component of $\rho(k,\omega_{\mu})$
defines the superfluid stiffness. In terms of the currents it reads
\begin{equation}
	\rho_0=\rho_{(k=0,\omega_{\mu}=0)}=\frac{1}{L^d L_{\tau}}
	\left\langle \left(\sum_{i,\tau}{\cal J}^x_{i,\tau}\right)^2\right
	\rangle=\frac{1}{L^{d-2} L_{\tau}}\langle w^2_x \rangle\,.
\end{equation}
The winding numbers $w_x=\frac{1}{L} \sum_{i,\tau} {\cal J}^x_{i,\tau}$
can be measured easily; they are only modified by periodic current
loops, local loops do not change their value. The symmetry in
$x/y$-direction further simplifies the numerics.

Direct measurement in the critical regime is ruled out by the
divergence of the correlation length $\xi$ at the continuous
transition. The powerful method of finite-size scaling, however,
takes advantage of this fact and allows us to study the critical
behavior.
Near a continuous phase transition the diverging coherence length
is cut off by the finite system size $L$. As a result all quantities
will depend on the ratio $\xi/L$. A finite-size scaling relation
for the superfluid stiffness $\rho_0$ is readily derived~\cite{cha}.
Assuming hyperscaling, the critical part of the free energy density
behaves as the inverse correlation volume $\xi^d\xi_{\tau}$. Here, $\xi$
is the correlation length in real space which may be different from
the correlation length $\xi_{\tau}$ in imaginary time, depending on
the quantum dynamics of the model. The dynamical critical exponent
$z$ can be introduced through $\xi_{\tau}\propto\xi^z$. Combining
Eq.~(\ref{df}), the hyperscaling relation and the fact that $\xi$
is cut off by $L$ ($\xi_{\tau}$ by $L_{\tau}$) we arrive at the
scaling behavior
\begin{equation}
	\rho_{0}={L^{2-d-z}}\tilde{\rho}
	(b L^{1/\nu}\delta, L_{\tau}/L^{z})\,.
\label{rhosc}
\end{equation}
Here, $b$ is a non-universal scale factor and $\tilde{\rho}$ is a
universal scaling function with a smooth dependence on $L_{\tau}/L^{z}$
and $\xi/L=\delta^{-\nu}/L$. We assumed a power law critical behavior
$\xi \propto \delta^{-\nu}$ with dimensionless distance to the
transition $\delta=(K-K^{\ast})/K^{\ast}$ and critical coupling
$K^{\ast}$. At the critical point ($\delta=0$),
$L^{d+z-2}\rho_{0}$ is a function of $L_{\tau}/L^{z}$ only. Thus, plots
of $L^{d+z-2}\rho_{0}$ vs.\ $K$ will intersect at the transition if
$L_{\tau}/L^{z}$ is kept constant, with no further fitting involved.
Furthermore, the data for $L^{d+z-2}\rho_{0}$ plotted as a function
of $L^{1/\nu}\delta$ for different system sizes should collapse onto
one single curve. This allows the exponent $\nu$ to be determined.

As we are interested also in diagonal LRO and a possible supersolid
phase, we measure the structure factor as well. The structure factor
is the $(\pi,\pi)$-component of the equal time density-density
correlation, see Eq.~(\ref{eqstr}), and may be derived as the second
variation of the free energy with respect to a staggered chemical
potential. Expressed in terms of the currents ${\cal J}^{\tau}$ it reads
\begin{equation}
	S_{\pi}=\frac{1}{L^{4}L_{\tau}}\Big\langle\sum_{ij,\tau}(-1)^{i+j}
	{\cal J}^{\tau}_{i,\tau}{\cal J}^{\tau}_{j,\tau}\Big\rangle\;.
\end{equation}
The scaling relation for the structure factor is
\begin{equation}
	S_{\pi}=L^{-2\beta/\nu}
	\tilde{S}(b'L^{1/\nu}\delta,L_{\tau}/L^{z})\;.
\end{equation}
This scaling relation arises, since the structure factor $S_{\pi}$
is related to the square of the staggered magnetization order parameter
$M=\sum_{i}(-1)^{i}{\cal J}^{\tau}_{i}$ for which the exponent $\beta$ is the
order parameter exponent. Near the phase transition,
$M\sim \delta^{\beta}$ and $\xi\sim \delta^{-\nu}$, from which
the quoted form follows. From a three-parameter fit to the scaling
relation for different system sizes, the exponents $\nu$ and $\beta$
as well as the critical coupling constant $K^{\ast}$ are determined.
Again, plots of $L^{2\beta/\nu}S_{\pi}$ will cross at the critical
point if the ratio $L_{\tau}/L^{z}$ is kept constant.

In both scaling relations a knowledge of $z$ is assumed. This is
well-known for the SI transition and the scaling of $\rho_0$, i.e.,
in general $z$=2, except for the tips of the lobes where $z$=1 as
dictated by particle hole-symmetry. This argument is based on a
coarse-graining treatment \cite{doni} that allows for the explicit
construction of a Ginzburg-Landau-Wilson free energy functional for
the superconducting order parameter \cite{fwgf,owfs,jk-cg,kz-cg},
from which $z$ can be read off.
For the transition related to the structure factor the situation
is less clear. We were, for instance, not yet successful in explicitly
constructing the Ginzburg-Landau-Wilson free energy functional for
the CDW transition. From symmetry arguments one expects $z$=1 at the
mean-field level~\cite{bafr}. Also the spin-wave analysis of
Ref.~\cite{gergy} predicts $z$=1. However, it is unclear whether or not
$z$ for the CDW will be renormalized by the coupling to the gapless
sound mode in the superfluid part. As a working hypothesis we take
$z$=1 and comment later on the consistency of this assumption.

\subsection{Numerical Results}
\label{subsec-nm}
We report about the SI transitions for on-site interaction and
nearest-neighbor interaction in the complete range of the chemical potential.
A phase diagram is mapped out and the scaling predictions are verified.
The existence of a supersolid phase in a finite region of the phase
diagram is firmly established.

The case of {\it on-site interaction} has been extensively studied in the
past years. A phase diagram consisting of lobe-like structures in the
$t-\mu$ or $J-n_{0}$ plane
is found. Mott-insulating lobes are centered around integer values of
$n_0$. This has been observed in mean-field approximations, $t/U$ expansions
\cite{fm-bh,fm-ex}, and in Monte Carlo simulations of the one-dimensional Bose
Hubbard model \cite{bsz}. The shape of the lobes in two dimensions is yet
unknown. Therefore we first map out this phase diagram which is shown in
Fig.~\ref{mottfig}.
We performed the simulations for fixed $n_0$ and tuned through the transition
by varying the effective coupling $K$. In the phase diagram in Fig.\
\ref{mottfig} this
corresponds to moving on horizontal lines through the phase transition.

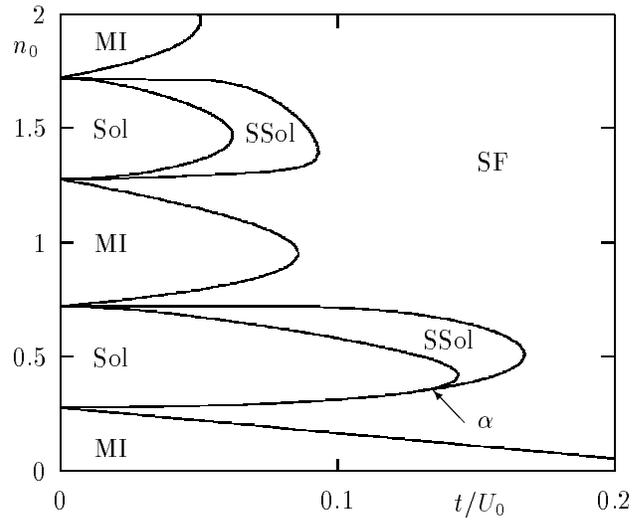
\begin{figure}
\caption{: Scaling plot of the Monte Carlo data for the superfluid stiffness
	$\rho_0$ for $n_0=0$ with on-site interaction only.}
\label{dat1fig}\end{figure}

At integer values of $n_0$ the system is particle-hole symmetric and we have
$z=1$. According to the scaling form of Eq.~(\ref{rhosc}) we use $L=L_{\tau}$
which keeps the second argument of the scaling function constant. Four
different system sizes with $L=6,8,10,12$ where studied. For the largest
systems we typically use $10^6$ sweeps through the system for equilibration
and measurement. In each sweep we try all possibilities of local and periodic
current loops. Without any fitting the curves of $\rho_0 L^{d+z-2}=\rho_0 L$
vs.\ $K$ should cross in the critical point $K^*$. This serves as a good test
of the scaling prediction. The critical exponent $\nu$ can be extracted by
fitting the data of different system sizes near the transition to one scaling
curve. This is shown in Fig.~\ref{dat1fig} where we plot the data as a
function of $\delta L^{1/\nu}$. For $n_0=0$ we obtain the critical coupling
$K^*=0.886\pm0.003$, and using a $\chi^2$ fit $\nu=0.69\pm0.06$. The exponent
$\nu$ agrees with the expected 3D-XY behavior, where $\nu\approx 2/3$.

Away from the symmetry points at the tips of the lobes the dynamical critical
exponent is $z=2$. In order to keep $L_{\tau}/L^z$ constant we simulate
systems
with $L_{\tau}=L^2/4$, i.e.,  $L\times L \times L_{\tau}=6 \times 6 \times 9$,
$8 \times 8 \times 16$ and $ 10\times 10\times 25$.
Now the curves of $\rho_0 L^2$ should cross in the critical point. We obtained
data for $n_0=$0.05, 0.1, ..., 0.45 as shown in Table \ref{table-ons}. From a
fit to the scaling form we find $0.42<\nu<0.52$ with errors of $\approx 20
\%$. This is in agreement with the expected mean-field transition where
$\nu=1/2$. A more accurate determination of the critical exponents turns out
to require an enormous numerical effort with our method and is beyond the
scope of our present work.

In Fig.~\ref{mottfig}, the Monte Carlo data are shown
together with mean-field phase boundaries and the results of the
$t/U$-expansion of Refs.~\cite{fm-bh,fm-ex} in the limit of large $n_{0}$ (or
$\mu$). A lobe shape is
observed, these lobes are sharper than predicted by mean-field, but smoother
than expected from the $t/U$-expansion. Approaching $n_0=0.5$ we observe that
the critical value of $J/U_0$ decreases and a simple extrapolation yields
$J^{\ast}=0$ for $n_0=0.5$. A numerical answer to the question whether
$J^{\ast}$ is zero at $n_0=0.5$ is not possible, since the acceptance rates
are very small for small $J/U_0$.

The inclusion of {\it nearest-neighbor interactions} yields a richer phase
diagram. The scaling for the superfluid stiffness is not modified by the
inclusion of short-range interactions, i.e., the universality class is
preserved. We expect $z=1$ in the case of particle-hole symmetry, $z=2$
otherwise. For nearest-neighbor interaction the lines with integer and
half-integer values of $n_0$ exhibit particle-hole symmetry. For the
transition related to charge-density wave order, we expect $z=1$ throughout
the phase diagram.
Guided by the
mean-field phase diagram we expect the existence of two different kinds of
lobes, with homogeneous particle densities centered around integer values of
$n_0$, and with a checkerboard charge-density wave centered around half integer
values of $n_0$. The supersolid phase is expected in the vicinity of the
checkerboard lobe.

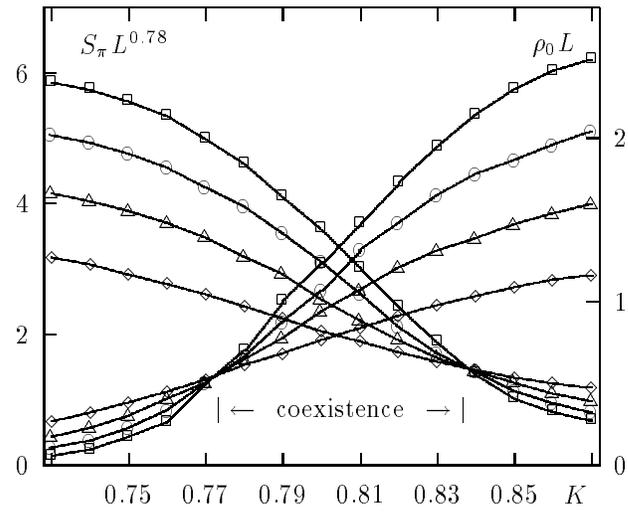
\begin{figure}
\caption{: Scaling plots for the structure factor $S_\pi$ for $z=1$ (lower
	curve) and $z=2$ (upper curve) at $n_{0}$= 0.4 and $U_1/U_0=1/5$.}
\label{fig:zdisc}\end{figure}

For $n_{0}$= 0 and 0.5 we simulated $L\times L\times L$ systems, where
$L$= 6, 8, 10, 12, as suggested by particle-hole symmetry and $z=1$.
For $n_{0}$=0.1, 0.2 and 0.4 we obtain $z=2$ for the superfluid transition.
In order to keep the ratio $L_{\tau}/L^{z}$ constant, we simulated
$L\times L\times L^{2}/4$ systems, where $L$= 6, 8, 10. For $n_{0}$=0.4 near
the transition for the diagonal LRO we performed the calculations with both
$z=1$ and $z=2$. The results are summarized in
Fig.~\ref{dat2fig} and Fig.~\ref{fig:zdisc} and Table \ref{table-nn}.
In the following we discuss our data in detail.

First we present our data for $n_{0}$=0.5. Figure \ref{dat2fig} shows that
there are two separate transitions for diagonal and off-diagonal LRO with a
coexistence region in between where {\it both} the superfluid density {\it
and} the structure factor scale to a finite value in the thermodynamic
limit. This demonstrates the coexistence of diagonal LRO and off diagonal LRO
for soft-core bosons with nearest-neighbor interaction in 2
dimensions. Table~\ref{table-nn} shows that the exponent $\nu$ is different
for the two transitions. For the transition related to superfluidity (point
$\beta$ in Fig.~\ref{cbssfig}) we find $\nu=0.65\pm0.08$ which is consistent
with the 3D XY universality class. For the transition related to crystalline
order (point $\gamma$) the universality class is not known. We find
$\nu\approx$0.55 and $\beta\approx$0.21.

Also at $n_{0}$=0.4 we find two separate transitions that are the boundaries
for the supersolid phase in between, see Table \ref{table-nn}. As compared to
$n_{0}$=0.5 both transitions are shifted to smaller values of the coupling
constant $K$. This is consistent with the mean-field phase diagram. Again the
two transitions have different critical exponents. The transition related to
superfluidity (the line separating phases ``Sol'' and ``SSol'' in
Fig.~\ref{cbssfig})
has $\nu=0.44\pm0.08$ which is consistent with a mean-field transition in
$d+z=4$ effective dimensions. For the transition related to crystalline order
(between phases ``SSol'' and ``SF'') we compare the data for $z=2$ and $z=1$ in
Fig.~\ref{fig:zdisc}.
A fit to the expected scaling behavior is equally possible in
both cases. The statistical errors of the data for $z=2$ are larger,
since the simulation for $z=2$ requires larger $L_{\tau}$ (up to 25) which
decreases the acceptance ratio for periodic loops in this direction.
The values for the critical coupling and the critical exponents for
both $z=1$ and $z=2$ are shown in Table \ref{table-nn}, they coincide
within the error bars. With the values of Table~\ref{table-nn} the scaling
relation $2\beta=\nu(d+z-2+\eta)$ leads to $0.8\pm0.2=z+\eta$, with
correlation function exponent $\eta$ (small and usually positive). This rules
out $z=2$ and one is led to conclude that indeed $z\sim 1$ for the crystalline
transition, independent of $n_{0}$.

The small value of $\beta$ rules out mean-field behavior for the CDW
transition. Insight into the nature of this transition is gained by the
following consideration. In the neighborhood of this transition, i.e., far in
the superfluid phase, the x,y-components of the currents ${\cal J}$ fluctuate
strongly and may be integrated out as Gaussian fluctuations. In other words:
one expects the staggered magnetization order parameter to couple to the
gapless 4$^{th}$ sound mode of the superfluid.
This yields strong long-range interactions in the time direction for the
$\tau$-components of the currents ${\cal J}$. It is likely that these
long-range interactions are a relevant perturbation and suppress the
exponent $\beta$.

Finally, we discuss the data for $n_{0}$ = 0, 0.1, and 0.2. In these cases
there is only one phase transition, as the Mott-insulating lobes
(phase ``MI'' in
Fig.~\ref{cbssfig}) do not have any non-trivial crystalline order. Our
data are consistent with a transition in the 3D XY universality class
for $n_{0}$=0 and a mean-field phase transition for $n_{0}$ = 0.1 and 0.2.

\section{Discussion}
\label{sec-d}
We studied the $T=0$ phase diagram of two-dimensional interacting
bosons by various techniques. The combination of the various aspects
of these treatments allow us to draw the following conclusions.

For on-site interaction we have seen the lobe structure of the phase boundary
in the $t-\mu$ or $J-n_{0}$ plane. This structure is exactly periodic in the
chemical potential $n_{0}$ in the Quantum-Phase model of Eq.~(\ref{qp}). In
mean
field or variational treatments of the different models, the lobes are
parabolic near their tips. The Monte Carlo results, however, show that the
lobes are sharper than predicted by mean-field approximations. This deviation
near the tips of the lobes is expected by scaling arguments: due to the
additional particle-hole symmetry at the tips of the lobes, the effective
dimension is reduced to $d_{eff}=d+z=3$ and enhanced fluctuations destroy part
of the ordered (superfluid) phase. This is consistent with the value of the
critical exponent $\nu$ obtained from the simulations ($\nu\approx 2/3$) that
agrees well with a 3D XY critical point. Similar sharpening of the lobes is
seen in a higher-order $t/U$ expansion for the BH model~\cite{fm-bh,fm-ex}.

A central point of this paper is the consideration of finite-range
interactions. They lead to a richer phase diagram, due to commensurability
effects. In particular, charge-density wave or solid order
appears in parts of the phase diagram.
In our definition the solid phase is characterized by the breaking
of the translation symmetry of the underlying lattice, i.e., if a
density wave with a wave vector smaller than the lattice is present.

Upon inclusion of nearest-neighbor interactions new half-filled lobes
around half-integer $n_{0}$ appear in the phase diagram, in which the
checkerboard ($\pi,\pi$) density wave is stable. For
next-nearest
neighbor interactions additional density waves exist in which 1 or 3 out of
4 sites are occupied, around $n_{0}$=1/4 or 3/4. These pure density
waves are incompressible Mott-insulators.
Surprisingly, each Mott-insulating charge-density wave seems to have a
corresponding compressible supersolid phase in which the particular
CDW order coexists with off diagonal LRO.

For soft-core bosons the supersolid phases appear already for nearest-neighbor
interaction. Hard-core bosons, on the other hand, supersolidify only if at
least next-nearest neighbor interaction is present. Another difference is the
presence of a supersolid at the tip of the half lobe: hard-core bosons only
form a supersolid away from half-integer filling, whereas soft-core bosons do
so also at half-integer filling.

The difference between Bose-Hubbard and Quantum-Phase models is another
issue. Our results indicate that, on the mean-field level, the two models are,
apart from a trivial rescaling of the hopping matrix element, almost identical
for large $\mu$ or $n_{0}$. For small $\mu$ corresponding to a filling of less
than one boson per site the mean-field solution for the two models differ
essentially and are not described by a simple rescaling of the
parameters. Within our mean-field approximation we find that the supersolid
phase is suppressed around the solid lobe with exactly half filling
(alternating empty and singly-occupied sites). This may be one of the reasons
why in a recent numerical simulation of the BH model \cite{gergy} a supersolid
phase was not found for exactly half filling.

Our combined mean-field and Monte Carlo analysis has shown that the
density-wave order transition is not mean-field like. Both the exponents
($\beta\approx 0.21$ and $\nu\approx 0.55$) and the location of the
phase boundary deviate considerably from the mean-field results
($\beta= 1/2$ and $\nu= 1/2$), see also Fig.~\ref{cbssfig}.
This may be related to the coupling of the CDW order parameter to the
gapless $4^{th}$ sound mode in the superfluid and remains a subject for
further study.
A related issue is the value of the dynamical critical exponent $z$
for this transition. The value $z$=1 used in the simulations gives good
scaling. However, other values of $z$ turn out to give reasonable
scaling as well and yield identical values for $\beta$, $\nu$ and
$K^{\ast}$.
It seems that the CDW transition is rather indifferent
to the value of $z$ used to scale the systems studied numerically.
 From the scaling relation
$2\beta/\nu=d+z-2+\eta$ we expect the combination $z$+$\eta\approx$
0.8$\pm$0.2 which would indicate either a small $\eta$ and renormalized
$z\approx$ 0.8 or $z\approx 1$ and $\eta$ negative. Further simulations
and renormalization group studies are required for solving this
problem.

In summary we performed an analysis of interacting bosons in 2D
within zero temperature lattice models, using and comparing both
mean-field theory and exact Quantum Monte Carlo simulations. We obtained
the full phase diagram for on-site interaction as well as for
nearest-neighbor interactions, in which case our simulations establish the
existence of a supersolid phase in which a CDW and superfluidity coexist.

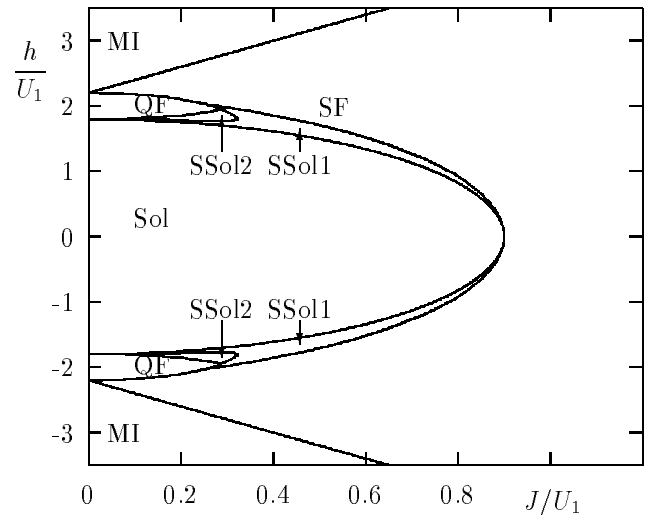
\begin{figure}
\caption{: Phase diagrams for soft-core bosons, as obtained from the
	mean-field analysis of the QP model with on-site interaction
	$U_{0}$ only
	(round curves on the left). The symbols	are the Monte Carlo data as
	discussed in Sect.~\protect\ref{sec-nm}. The sharp lobes are results
	of the third-order $t/U$-expansion of Ref.~\protect\cite{fm-bh}
	(extreme right) and the extrapolation to infinite order in $t/U$ from
	Ref.~\protect\cite{fm-ex}. The superfluid phase is denoted by ``SF''
	and the Mott-insulating phase by ``MI''.}
\label{mottfig}\end{figure}

{\bf Acknowledgments:} We thank G.~T. Zimanyi for motivation and
discussions, and J.~K. Freericks and H. Monien for discussions and their data
as plotted in Fig.~\ref{mottfig}. The work was partially supported by
'Sonderforschungsbereich 195' of the DFG and by the Swiss Nationalfonds (AvO).

\begin{table}
$$
\begin{array}{||c|c|c|c|c||c|c|c|c|c||}
\hline\hline
 n_{0}	& z 	& K^{\ast} 	& J^{\ast}/U_0	& \nu 	&
 n_{0} 	& z 	& K^{\ast}      & J^{\ast}/U_0	& \nu	\\ \hline\hline
 0.00	& 1	& 0.886\pm0.003	& 0.152		& 0.69\pm0.06 &
 0.25	& 2	& 0.64\pm0.01	& 0.098		& 0.5\pm0.1 	\\ \hline
 0.05	& 2	& 0.85\pm0.01	& 0.144		& 0.5\pm0.1 &
 0.30	& 2	& 0.56\pm0.01	& 0.081		&  ``	\\ \hline
 0.10	& 2	& 0.82\pm0.02	& 0.137		& ``	&
 0.35	& 2	& 0.47\pm0.01	& 0.062		&  ``	\\ \hline
 0.15	& 2	& 0.77\pm0.02	& 0.126		& ``	&
 0.40	& 2	& 0.37\pm0.01	& 0.042		&  ``	\\ \hline
 0.20	& 2	& 0.71\pm0.01	& 0.113		& ``	&
 0.45	& 2	& 0.26\pm0.01	& 0.023		&  ``	\\ \hline
\end{array}
$$
\caption{Critical couplings and exponents for on-site interactions}
\label{table-ons}
\end{table}

\begin{table}
$$
\begin{array}{||c||c|c|c|c||c|c|c|c|c||}
\hline\hline
\multicolumn{10}{||c||}{\mbox{the transition for}}
\\ \hline
\multicolumn{1}{||c||}{}&\multicolumn{4}{|c||}{\mbox{off-diagonal LRO}}&
\multicolumn{5}{|c||}{\mbox{diagonal LRO}}
\\ \hline
 n_{0}	& z 	& K^{\ast} 	& J^{\ast}/U_0	& \nu 	&
 	  z 	& K^{\ast}      & J^{\ast}/U_0	& \nu 	& \beta \\ \hline\hline
 0.5	& 1 	& 0.775\pm0.005 & 0.127		& 0.65\pm0.08 &
	  1 	& \; 0.837\pm0.005& 0.141 	& \;0.55\pm0.05 &
0.21\pm 0.04 \\ \hline
 0.4	& 2 	& 0.645\pm0.008 & 0.099		& 0.44\pm0.08 &
	  2 	& 0.749\pm0.006 & 0.122		& 0.5\pm0.1   & \; 0.25
\pm 0.10 \\ \hline
 0.4	&   	&		&      		&		&
	  1 	& 0.747\pm0.007 & 0.121		& 0.55\pm0.06	& \; 0.21
\pm 0.08 \\ \hline \hline
 0.2	& 2 	& 0.446\pm0.005 & 0.057		& 0.5\pm0.1   &
\multicolumn{5}{|c||}{\mbox{for comparison:}} \\ \hline
0.1	& 2 	& 0.707\pm0.007 & 0.112		&\; 0.49\pm0.11 &
\multicolumn{3}{|c|}{\;\;\;\; \mbox{mean-field:}}  &  1/2 & 1/2 \\ \hline
0.0	& 1 	& \; 0.843\pm0.005& 0.142		&  0.61\pm0.08 &
\multicolumn{3}{|c|}{\;\;\;\;\;\;\;\;\;\; \mbox{3D XY: }} & 2/3   &  1/3 \\
\hline\hline
\end{array}
$$
\caption{Critical couplings and exponents for the different
transitions with nearest-neighbor interactions $U_1/U_0=0.2$}
\label{table-nn}
\end{table}

\begin{figure}
\caption{: Phase diagram for hard-core bosons, as obtained from spin
	mean-field theory on the XXZ model. Nearest and
	next-nearest neighbor ($U_{2}/U_{1}=0.1$) interactions are included.
	Several phases appear: superfluid (SF), Mott-insulating (MI),
	checkerboard solid (Sol) and the corresponding supersolid (SSol1),
	and quarter-filled (QF) solid with the corresponding
	supersolid (SSol2).}
\label{spinfig}\end{figure}

\begin{figure}
\caption{: Phase diagrams for soft-core bosons, as obtained from the
	mean-field analysis of the QP model with on-site ($U_{0}$) and
	nearest-neighbor ($U_{1}/U_{0}=1/5$) interaction. The symbols are the
	Monte Carlo data as discussed in Sect.~\protect\ref{sec-nm}. The
	checkerboard charge-density wave is denoted by ``Sol'', the supersolid
	phase by ''SSol'', the superfluid phase is denoted by ``SF''
	and the Mott-insulating phase by ``MI''.}
\label{cbssfig}\end{figure}

\begin{figure}
\caption{: Mean-field phase diagram for nearest-neighbor and next-nearest
	neighbor interactions, $U_1/U_0=0.2$, $U_2/U_0=0.02$. In addition to
	the phase diagram in Fig.~\protect\ref{cbssfig}, small lobes at
	quarter filling and tree-quarter filling appear at small $J$ with their
	corresponding supersolids. The phases are labeled as in
	Fig.~\protect\ref{spinfig}
	.}
\label{nnnfig}\end{figure}

\begin{figure}
\caption{: Phase diagram for the Bose-Hubbard model as determined from the
	variational Ansatz with nearest-neighbor interactions, $U_1/U_0=1/5$.
	For small $n_{0}$ the supersolid phase vanishes at point $\alpha$.}
\label{bhvarfig}\end{figure}

\begin{figure}
\caption{: Monte Carlo data for the superfluid stiffness $\rho_0$ and the
	structure factor $S_{\pi}$ for $n_0=0.5$ with nearest-neighbor
	interaction $U_1/U_0=1/5$. We clearly observe two distinct transitions
	at the crossing points of these scaling plots, and a coexistence phase
	in between.}
\label{dat2fig}\end{figure}

\begin{references}
\bibitem{gold}  B.~G. Orr, H.~M. Jaeger, A.~M. Goldman, and C.~G. Kuper,
		Phys. Rev. Lett. {\bf 56}, 378 (1986);
		H.~M. Jaeger, D.~B. Haviland, B.~G. Orr, and A.~M. Goldman,
		Phys. Rev. B {\bf 40}, 182 (1989);
		D.~B. Haviland, Y. Liu, and A.~M. Goldman,
		Phys. Rev. Lett. {\bf 62}, 2180 (1989);
		Y. Liu, K.~A. Greer, B. Nease, D.~B. Haviland, G. Martinez,
		J.~W. Haley, and A.~M. Goldman,
		Phys. Rev. Lett. {\bf 67}, 2068 (1991).
\bibitem{hpsi}  A.~F. Hebard and M.~A. Paalanen,
		Phys. Rev. Lett. {\bf 65}, 927 (1990).
\bibitem{jja}	L.~J. Geerligs, M. Peters, L.~E.~M. de Groot, A. Verbruggen,
		and J.~E. Mooij, Phys. Rev. Lett. {\bf 63}, 326 (1989);
		H.~S.~J. van der Zant, F.~C. Fritschy, W.~E. Elion,
		L.~J. Geerligs,
		and J.~E. Mooij, Phys. Rev. Lett. {\bf 69}, 2971 (1992);
		H.~S.~J. van der Zant, L.~J. Geerligs, and J.~E. Mooij,
		Europhys. Lett. {\bf 19}, 541 (1992);
		T.~S. Tighe, M.~T. Tuominen, J.~M. Hergenrother,
		and M. Tinkham, Phys. Rev. B {\bf 47}, 1145 (1993);
		P. Delsing, C.~D. Chen, D.~B. Haviland, Y. Harada,
		and T. Claeson, Phys. Rev. B {\bf 50}, 3959 (1994);
		H.~S.~J. van der Zant, W.~J. Elion, L.~J. Geerligs,
		and J.~E. Mooij, unpublished.
\bibitem{efe}   K.~B. Efetov, Sov. Phys. JETP {\bf 51}, 1015 (1980).
\bibitem{doni}	S. Doniach, Phys. Rev. B {\bf 24}, 5063 (1981).
\bibitem{fwgf}  M.~P.~A. Fisher, B.~P. Weichman, G. Grinstein,
		and D.~S. Fisher, Phys. Rev. B {\bf 40}, 546 (1989).
\bibitem{f-db}	M.~P.~A. Fisher, Phys. Rev. Lett. {\bf 65}, 923 (1990).
\bibitem{zimrev}G.~T. Zimanyi in {\em Strongly Correlated Electronic
		Materials}, edited by K.~S. Bedell {\it et al.}
		(Addison-Wesley, 1994).
\bibitem{andlif}A.~F. Andreev and I.~M. Lifshitz,
		Sov. Phys. JETP {\bf 29}, 1107 (1969).
\bibitem{l-ss}	A.~J. Leggett, Phys. Rev. Lett. {\bf 25}, 1543 (1970).
\bibitem{mt-ss}	H. Matsuda and T. Tsuneto,
		Suppl. Prog. Theor. Phys {\bf 46}, 411 (1970).
\bibitem{lf-ss}	K. Liu and M. Fisher,
		J. Low. Temp. Phys. {\bf 10}, 655 (1973).
\bibitem{bfs}	C. Bruder, R. Fazio, and G. Sch\"{o}n,
		Phys. Rev. B {\bf 47}, 342 (1993);
		C. Bruder and G. Sch\"on, in {\em KT Transition and
		Superconducting Arrays}, edited by D. Kim {\it et al.},
		(Min Eum Sa Co., Seoul, 1993), p. 175.
\bibitem{rs-ss} E. Roddick and D.~H. Stroud,
		Phys. Rev. B {\bf 48}, 16600 (1993).
\bibitem{m-ss}  M.~W. Meisel, Physica {\bf 178}, 121 (1992);
		and references therein.
\bibitem{lg-ss} G.~A. Lengua and J.~M. Goodkind,
		J. Low. Temp. Phys. {\bf 79}, 251 (1990).
\bibitem{vavl}	M. Gabay and A. Kapitulnik,
		Phys. Rev. Lett. {\bf 71}, 2138 (1993);
		S.~C. Zhang, Phys. Rev. Lett. {\bf 71}, 2142 (1993).
\bibitem{hex}	K. Mullen, H.~T.~C. Stoof, M. Wallin, and S.~M. Girvin,
		Phys. Rev. Lett. {\bf 72}, 4013 (1994).
\bibitem{leon}  L. Balents, unpublished.
\bibitem{gergy} G.~G. Batrouni, R.~T. Scalettar, G.~T. Zimanyi,
		and A.~P. Kampf, Phys. Rev. Lett. {\bf 74}, 2527 (1995);
		A.~P. Kampf, G.~T. Zimanyi, G.~G. Batrouni, and
		R.~T. Scalettar, unpublished.
\bibitem{rs-mc}	E. Roddick and D.~H. Stroud,
		Phys. Rev. B {\bf 51}, 8672 (1995).
\bibitem{nelson}D.~R. Nelson, Phys. Rev. Lett. {\bf 60}, 1973 (1988).
\bibitem{fl-d}  M.~P.~A. Fisher and D-H. Lee,
		Phys. Rev. B {\bf 39}, 2756 (1989).
\bibitem{feigel}M.~V. Feigel'man, V.~B. Geshkenbein, L.~B. Ioffe,
		and A.~I. Larkin, Phys. Rev. B {\bf 48}, 16641 (1993).
\bibitem{blatt} G. Blatter, M.~V. Feigel'man, V.~B. Geshkenbein,
		A.~I. Larkin, and V.~M. Vinokur,
		Rev. Mod. Phys. {\bf 66}, 1125 (1994).
\bibitem{frey}  E. Frey, D.~R. Nelson, and D.~S. Fisher,
		Phys. Rev. B {\bf 49}, 9723 (1994).
\bibitem{baltin}R. Baltin, K-H. Wagenblast, G. Sch\"{o}n,
		and A. van Otterlo, unpublished.
\bibitem{ow-ss} A. van Otterlo and K-H. Wagenblast,
		Phys. Rev. Lett. {\bf 72}, 3598 (1994).
\bibitem{bfkos} C. Bruder, R. Fazio, A.~P. Kampf, A. van Otterlo, and
		G. Sch\"{o}n, Physica Scripta T {\bf 42}, 159 (1992).
\bibitem{skpr}	K. Sheshadri, H.~R. Krishnamurthy, R. Pandit, and T.~V.
		Ramakrishnan, Europhys. Lett. {\bf 22}, 257 (1993).
\bibitem{rmwh}  N.~D. Mermin and H. Wagner,
		Phys. Rev. Lett. {\bf 17}, 1133 (1966);
		P.~C. Hohenberg, Phys. Rev. {\bf 158}, 383 (1967).
\bibitem{khw-d} K-H. Wagenblast,
		Diplomarbeit Universit\"{a}t Karlsruhe 1994, unpublished.
\bibitem{gutz}  M. Gutzwiller, Phys. Rev. Lett. {\bf 10}, 159 (1963).
\bibitem{huse}  D.~A. Huse, and V. Elser,
		Phys. Rev. Lett. {\bf 60}, 2531 (1988).
\bibitem{roko}	D.~S. Rokhsar, and B.~G. Kotliar,
		Phys. Rev. B {\bf 44}, 10328 (1991).
\bibitem{kraut}	W. Krauth, M. Caffarel, and J.-P. Bouchard,
		Phys. Rev. B {\bf 45}, 3137 (1992).
\bibitem{bsz}   G.~G. Batrouni, R.~T. Scalettar, and G.~T. Zimanyi,
		Phys. Rev. Lett. {\bf 65}, 1765 (1990);
		R.~T. Scalettar, G.~G. Batrouni, and G.~T. Zimanyi,
		Phys. Rev. Lett. {\bf 66}, 3144 (1991).
\bibitem{nfsb}  P. Niyaz, C.~Y. Fong, R.~T. Scalettar, and G.~G. Batrouni,
		Phys. Rev. B {\bf 50}, 362 (1994);
		P. Niyaz, R.~T. Scalettar, C.~Y. Fong, and G.~G. Batrouni,
		Phys. Rev. B {\bf 44}, 7143 (1991).
\bibitem{kt-mc} W. Krauth and N. Trivedi,
		Europhys. Lett. {\bf 14}, 627 (1991).
\bibitem{swgy}  E.~S. S\o rensen, M. Wallin, S.~M. Girvin and A.~P. Young,
		Phys. Rev. Lett {\bf 69}, 828 (1992);
		M. Wallin, E.~S. S\o rensen, S.~M. Girvin, and A.~P. Young,
		Phys. Rev. B {\bf 49}, 12115 (1994).
\bibitem{fm-bh} J.~K. Freericks and H. Monien,
		Europhys. Lett. {\bf 26}, 545 (1994).
\bibitem{fm-ex} J.~K. Freericks and H. Monien, unpublished.
\bibitem{jkkn}  J.~V. Jos\'{e}, L.~P. Kadanoff, S. Kirkpatrick,
		and D.~R. Nelson, Phys. Rev. B {\bf 16}, 1217 (1977);
		L.~P. Kadanoff, J. Phys. A {\bf 11}, 1399 (1978).
\bibitem{vill}  J. Villain, J. Physique {\bf 36}, 581 (1975).
\bibitem{cha}	M.-C. Cha, M.~P.~A. Fisher, S.~M. Girvin, M. Wallin,
		and A.~P. Young, Phys. Rev. B {\bf 44}, 6883 (1991);
		M.~P.~A. Fisher, G. Grinstein, and S.~M. Girvin,
		Phys. Rev. Lett. {\bf 64}, 587 (1990).
\bibitem{owfs}  A. van Otterlo, K-H. Wagenblast, R. Fazio, and G. Sch\"{o}n,
		Phys. Rev. B {\bf 48}, 3316 (1993).
\bibitem{jk-cg}	J.~G. Kissner and U. Eckern, Z. Phys. B {\bf 91}, 155 (1993).
\bibitem{kz-cg}	A.~P. Kampf and G.~T. Zimanyi,
		Phys. Rev. B {\bf 47}, 279 (1993).
\bibitem{bafr}  L. Balents, and E. Frey, private communication.
\end{references}
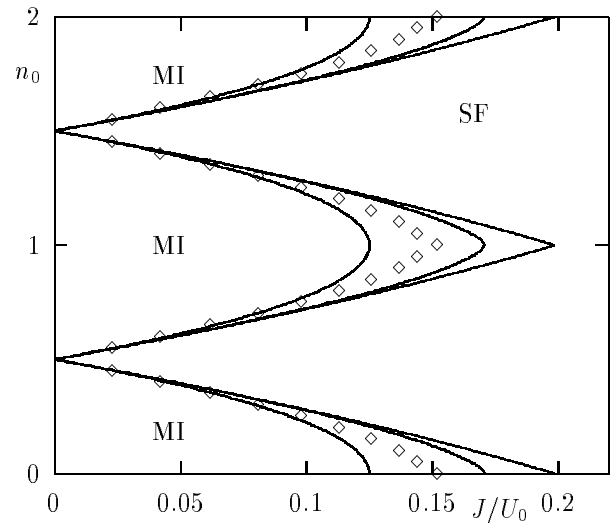
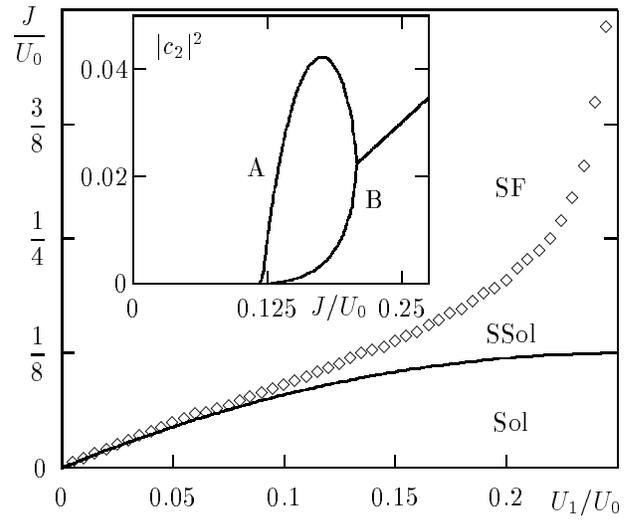
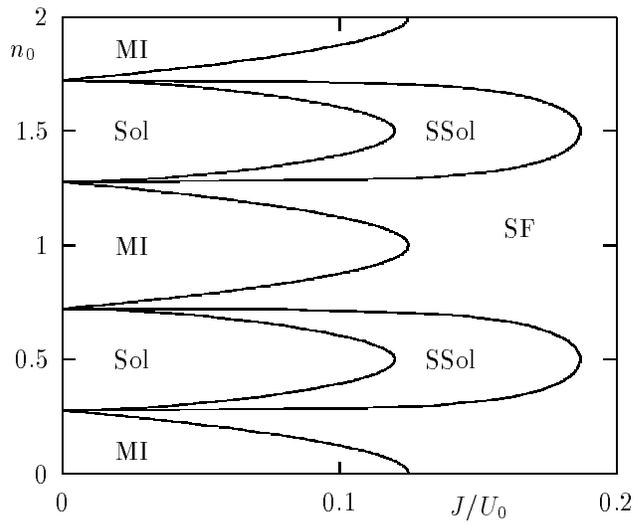
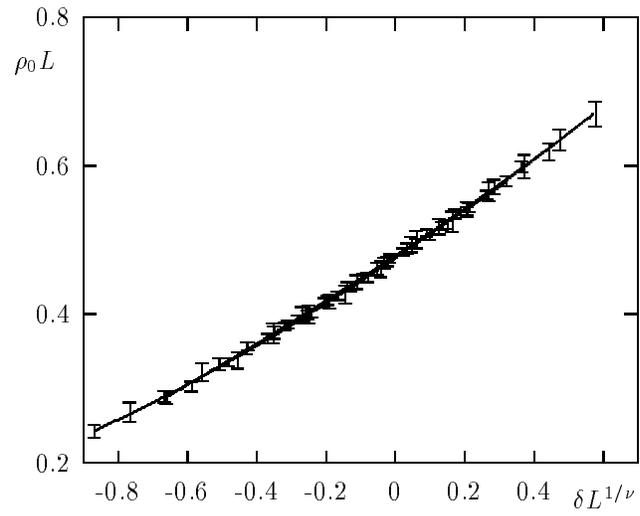
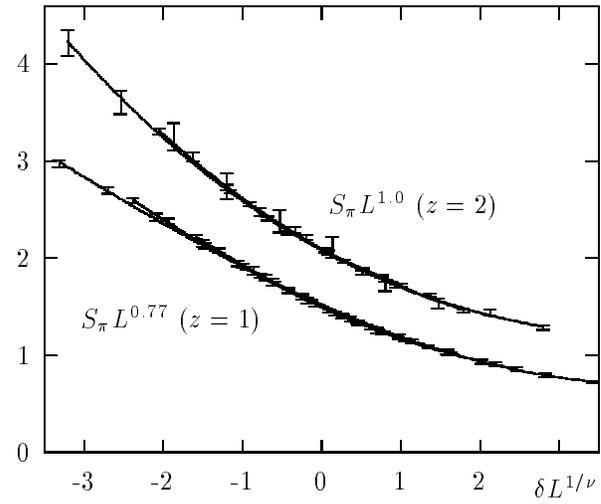
\end{document}